\documentclass[aps,pre,showpacs,amsfonts,amssymb,amsmath,floatfix,onecolumn,
floats,nobalancelastpage,endfloats*,byrevtex,preprint]{revtex4}

\usepackage{graphicx}

\begin{document}


\title{MCAMC: An Advanced Algorithm for Kinetic Monte Carlo Simulations: 
from Magnetization Switching to Protein Folding}

\author{M.~A.\ Novotny and Shannon M.\ Wheeler}
\email{man40@ra.msstate.edu}
\affiliation{Department of Physics and Astronomy, and the MSU ERC, 
P.O. Box 5167, Mississippi State, MS 39762-5167}

\date{\today}

\begin{abstract}
We present the Monte Carlo with Absorbing Markov Chains 
(MCAMC) method for extremely long kinetic Monte Carlo simulations.  
The MCAMC algorithm does not modify the system dynamics.  
It is extremely useful for models with discrete 
state spaces when low-temperature simulations are desired.  
To illustrate the strengths and limitations of this algorithm we 
introduce a simple model involving random walkers on an energy 
landscape.  This simple model has some of the characteristics of 
protein folding and could also be experimentally realizable in 
domain motion in nanoscale magnets.  We find that even the 
simplest MCAMC algorithm can speed up calculations by many orders 
of magnitude.  More complicated MCAMC simulations can gain 
further increases in speed by orders of magnitude.  
\end{abstract}

\maketitle

\section{Introduction}

There are many excellent algorithms to decrease the computer 
time required to perform Monte Carlo simulations for the statics 
of a model system.  For example, see the articles in this volume 
by D.P.\ Landau and by B.\ D{\"u}nweg.  These algorithms and most other 
acceleration algorithms change the underlying dynamics of the 
system, which is permissible if and only if just the statics of the 
model is of physical interest.  

However, sometimes the dynamics of the simulation is 
physically relevant.  One example is metastability in 
the Ising model \cite{RikGor,MANmeta}.  
By coupling a lattice of quantum spin $\frac{1}{2}$ particles to 
a quantum heat bath it is possible to obtain the time-dependent density 
matrix of the quantum system of particles plus the bath.  
If the mixing of the quantum bath is much faster than the relaxation 
of the spins (and using a few other assumptions), the quantum bath can be 
integrated over to obtain a time-dependent Master equation 
\cite{Glau63,BinSto73,Bin74} for a 
classical spin $\frac{1}{2}$ Ising model on the same lattice.  
The transition rates in the Master equation are related to physical 
constants such as wave velocities as well as to expectation values 
in the original quantum system.  
This calculation was performed with a particular fermionic bath 
by Martin in 1977 \cite{Mart77}, and the Glauber dynamic \cite{Glau63} 
was obtained.  
Recently, Park et al.\ \cite{Park1,Park2} performed the same 
calculation with a $d$-dimensional bosonic bath and obtained 
somewhat different transition rates in the Master equation.  
This bosonic dynamic is relevant for molecular magnets, and also 
leads to some novel features in metastable decay \cite{Park2}.  
The dynamic obtained in the above fashion is physical, 
and hence cannot be modified if the time-dependence of the 
model is to be compared with experiments.  Note that this 
approach relates the dynamic Monte Carlo simulation time, 
measured in Monte Carlo Steps per Spin (MCSS), to 
the physical time (in seconds).  

For the Ising simulation, the attempt frequency in the kinetic Monte Carlo 
is related to the inverse phonon frequency, about $10^{-13}$~s.  
To study metastable decay the time scales are typically 
on the order of human times (seconds to years), 
or for relevance to paleomagnetism the time scale is many millions of years.  
Consequently, algorithms which {\it do not change the dynamic\/} 
but are {\it faster-than-real-time\/} must be used to directly 
compare with experiments.  This paper details one such algorithm, 
the Monte Carlo with Absorbing Markov Chains (MCAMC) algorithm 
\cite{MCAMC1,MCAMC2}.  
A recent review at the introductory level of faster-than-real-time 
kinetic Monte Carlo algorithms, including the MCAMC algorithm, 
is available \cite{MCReview}.  

\section{Model}

MCAMC simulations for the Ising model have been presented 
previously \cite{MCAMC1,MCAMC2,MCReview,MANTucson}.  
They are related to magnetization switching in thin 
nanoscale highly-anisotropic magnetic films.  
In this paper we 
introduce a simple model to illustrate the MCAMC method.  
We will see below that this simple model also has some 
interesting physics, and also can be related both to 
magnetization switching of thin films and to 
questions related to protein folding.  

Consider a one-dimensional lattice where each site $i$ has 
been assigned an energy $E_i$.  
This model is generalizable to higher dimensions, and the MCAMC 
method will work in higher dimensions, but here for 
simplicity we focus on the linear lattice.  
In particular, consider the 20-site lattice shown in 
Fig.~1, with energies given in Table~1.  

On this lattice we initially 
randomly place a number of walkers, $N_{\rm w}$.  
Introduce the dynamic that at each Monte Carlo step 
a walker is randomly picked (with uniform probability), 
and then whether the walker attempts to move left or right 
is randomly chosen.  
The chosen walker then moves to the adjacent lattice site with 
a probability 
\begin{equation}
\label{pmove}
p_{\rm move} =
\frac{{\exp(-E_{i\pm1}/k_{\rm B}T)}}
{{\exp(-E_i/k_{\rm B}T)+\exp(-E_{i\pm1}/k_{\rm B}T)}}
\; ,
\end{equation}
where the $+$ ($-$) sign is for the attempted move to the right (left).
We place reflecting walls (infinite $E$) at sites~0 and 21.  
One attempted move is a Monte Carlo step (mcs), and $N_{\rm w}$ 
attempts is a Monte Carlo step per walker (MCSW).  
We will be interested in the average time, $\langle\tau\rangle$, 
for all $N_{\rm w}$ walkers to first reach the same lattice site.  
This `coagulation' of walkers may or may not be at the 
global energy minima (site~9).  

\begin{table}[htbp]
\caption{The 20 energies used in the simulation, as in Fig.~1.}
\begin{tabular}{r | r}\\
\hline
{} Site Number {} & Energy \\
\hline
  1 &  1.0  \\
  2 &  1.0  \\
  3 &  0.0  \\
  4 &  0.5  \\
  5 &  0.0  \\
  6 &  0.0  \\
  7 &  0.5  \\
  8 &  0.0  \\
  9 & -2.0  \\
 10 &  0.5  \\
 11 &  0.5  \\
 12 &  0.75 \\
 13 &  0.0  \\
 14 &  0.5  \\
 15 &  0.0  \\
 16 & -1.0  \\
 17 & -1.0  \\
 18 &  0.0  \\
 19 &  0.0  \\
 20 & -1.0  \\
\hline\\
\end{tabular}
\end{table}

This model can be viewed as a model for switching in a nanomagnet 
where a domain wall is constrained to lie in a thin film 
that is anchored at the two ends by macroscopic magnets.  
The energy at each site would then correspond to the 
energy the domain wall would have when it is in a particular 
coarse-grained location in the thin film.  
The lowest energy points would be where pinning of the 
magnetic domain wall is the strongest and the highest energies 
would be the saddle points a domain wall would need to 
thermally overcome to traverse from one pinning site to another.  
The physical question is when $N_{\rm w}$ independent 
thin-film devices which start in a random state (demagnetized) 
would all first be in the same location.  

The same model can be viewed as a very simple example for 
protein folding by considering each point to be the free-energy 
at some abstract point in phase space \cite{Scrip}.  
Then the lowest energy corresponds to the native configuration.  
The physical question would then be the average speed at which the 
protein folds, i.e.\ when $N_{\rm w}$ proteins would be in the 
same configuration.  

\section{Kinetic Monte Carlo}

A number, $M$, of different coagulations of walkers will be performed to 
obtain the average lifetime $\langle\tau\rangle$, i.e.\ the 
average time until coagulation occurs.  
For each coagulation the $N_{\rm w}$ walkers are first placed randomly, 
with uniform probability on each site.  Then the kinetic Monte Carlo 
simulation is performed, calculating the number of time steps $\tau_i$ 
for coagulation $i$ until all $N_{\rm w}$ walkers are simultaneously at 
the same lattice site.  This lattice site may be any of the 20 sites 
of the lattice.  The average lifetime is then given by 
\begin{equation}
\langle\tau\rangle = \frac{1}{M} \sum_{i=1}^M \tau_i .
\end{equation}

The kinetic Monte Carlo algorithm is simple.  
Each move requires 3 uniformly distributed random numbers, 
$r_i$, with $0<r_i<1$.  
The first random number is used to select uniformly 
one of the $N_{\rm w}$ walkers.  This is accomplished by choosing 
walker $j$, given by $j=1+\lfloor r_1 N_{\rm w}\rfloor$ where 
$\lfloor x \rfloor$ is the integer part of $x$.  
If $r_2\le\frac{1}{2}$, the chosen walker attempts a move to the 
left, otherwise it attempts a move to the right.  
Finally if $r_3\le p_{\rm move}$ with $p_{\rm move}$ 
from Eq.~(\ref{pmove}) the walker moves to the chosen 
new lattice position.  Whether or not a move is made, the time 
is advanced by one unit,  $\tau_i = \tau_i + 1$.  

The results for $\langle\tau\rangle$ using this algorithm 
are shown in Fig.~\ref{fig:f2} and \ref{fig:f3}.  The data 
will be discussed in the results section, but here 
and in the next two sections we concentrate on 
the algorithmic aspects.  The CPU (central processing unit) time 
required (on a single processor of a Cray SV1 vector computer) 
for this algorithm is shown in Fig.~\ref{fig:f4} with 
the label {\it mc1}.  Note that at both very low and very high 
temperatures this algorithm requires substantial amounts of computer 
time.  In particular, the average CPU time required for this algorithm is 
proportional to the value of $\langle\tau\rangle$.  Thus at low 
temperatures where $\langle\tau\rangle$ grows exponentially fast in 
$T^{-1}$ the time required for the simulation also grows exponentially 
quickly.  

\section{$n$-fold way = $s$$=$$1$ MCAMC}

To decrease the simulation time required, without altering the 
dynamic, an event-driven simulation can be performed.  
This is also called an $n$-fold way simulation \cite{nfoldBKL}.  
The original paper \cite{nfoldBKL} used continuous time, but the 
same algorithm can be cast into the discrete-time version 
(where $\tau_i=\tau_i+1$ at each time step) 
\cite{nfoldMark} used here.  
The $n$-fold way algorithm is a 
$s$$=$$1$ MCAMC algorithm, because the system has 
one transient state ($s$$=$$1$), which is the current state 
of the system.  In our case it has $2 N_{\rm w}$ absorbing states 
(states the system can jump to from the current state), two 
(one of which may have zero probability) for 
each random walker.  

The $n$-fold way algorithm also requires 3 random numbers, $r_i$ 
at each step.  
First form a vector ${\vec p}_{\rm jump}$ of length 
$N_{\rm w}$ which contains the $N_{\rm w}$ probabilities 
$p_{\rm jump}(k)$ that 
a walker jumps to either the left or right {\it given\/} that it 
was picked in the uniform picking part of the algorithm.  
Then increment the time (in mcs) by 
\begin{equation}
\tau_i = \tau_i + \left\lfloor
\frac{\ln(r_1)}{\ln\left(1-\frac{p_{\rm sum}}{N_{\rm w}}\right)}
\right\rfloor + 1
\end{equation}
with $p_{\rm sum}=\sum_{k=1}^{N_{\rm w}} p_{\rm jump}(k)$.  
Next the walker, $j$, that actually jumped is calculated by finding 
the value of $j$ that satisfies 
\begin{equation}
\sum_{k=1}^{j-1} p_{\rm jump}(k) \le r_2 p_{\rm sum} < 
\sum_{k=1}^{j} p_{\rm jump}(k)
\label{jpick}
\end{equation}
where the first sum is taken as zero if $j=1$.  
Finally if the probability of moving left, $p_{\rm move,-}$, from 
Eq.~(\ref{pmove}) satisfies the relation
\begin{equation}
p_{\rm move,-} \le r_3 p_{\rm jump}(j)
\end{equation}
the $j^{\rm th}$ walker is moved to the left, otherwise it is moved 
to the right.  
This algorithm is repeated until all walkers coagulate at one point.  

The results for this discrete-time $n$-fold way algorithm is 
identical (within statistical errors for these $M=10^3$ coagulations) 
to those of the standard kinetic Monte Carlo.  The way the dynamic 
has been implemented on the computer is just different.  
Results from this algorithm are labeled {\it nf1\/} in the figures.  
The $n$-fold way algorithm 
requires additional calculations at each step, but the time increment 
that is added to $\tau_i$ may be larger than unity at each step.  
This can drastically decrease the simulation time required, particularly 
at low temperatures, as seen in Fig.~\ref{fig:f4}.  This is because the 
$n$-fold way algorithm is an event-driven algorithm.  In other words the 
time is incremented only when an event happens, namely only when a 
walker jumps from its current site.  The number of time steps 
before one walker jumps can be very large, particularly at low temperatures.  
As seen in Fig.~\ref{fig:f4} the $n$-fold way algorithm requires 
about an order of magnitude less CPU time at low temperatures than 
the previous implementation of the algorithm.  A factor of 10 is 
important in simulations, but at low temperatures where 
$\langle\tau\rangle$ is growing exponentially with $T^{-1}$, so does the 
required simulation time using the $n$-fold way method.  

\section{MCAMC with $s$$=$$2$}

At low temperatures it would be very nice to have a faster algorithm 
than the $n$-fold way.  The reason the $n$-fold way algorithm scales 
so poorly (exponentially in $T^{-1}$), is because when a walker 
is in a flat energy minima (at sites 5 and 6 or at sites 16 and 17) 
the average time before the walker moves is equal to a value which 
is independent of temperature.  In particular, given that the 
walker in one of these minima is picked it has a probability of 
$\frac{1}{4}$ of moving to the adjacent equal-energy site.  Thus at 
low temperatures the walker in such a minima will rattle back and 
forth many times before it jumps.  

To user higher $s$ MCAMC, more states are included in the transient 
subspace \cite{MCAMCMark}.  One way of doing this for a model 
closely related to the current model is given in \cite{MCReview}.  
That way is easily generalized to cases where the energies in 
the minima are nearly equal, or to the case of including 
larger numbers of transient states in the calculation.  Here we introduce 
a simple method that works most easily when the energies in some 
minima are equal and the energies to hope from these minima are equal.  

Let $N_{s1}$ be the number of walkers that are not located at 
sites 5-6 or 16-17, and $N_{s2}$ the number that are in one of these 
minima.  Clearly $N_{\rm w}=N_{s1}+N_{s2}$.  If all the walkers are located 
in either the 5-6 or the 16-17 minima, the $n$-fold way algorithm of the 
previous section must be used.  Otherwise, consider the current state 
of the system to be expanded to be the state with the 
$N_{s1}$ walkers fixed at their current site and the $N_{s2}$ 
walkers are still located in their respective 5-6 or 16-17 minima.  
Form the vector of length $N_{\rm w}$ with elements either: 
1) if the walker is not at the 5-6 or 16-17 minima, 
the ($n$-fold way) probability 
$p_{\rm jump}(k)$ that 
a walker jumps to either the left or right {\it given\/} that it 
was picked in the uniform picking part of the algorithm; 
2) if the walker is at the 5-6 or 16-17 minima, the probability 
that the walker exits this minima {\it given\/} that it was picked 
during the random picking part of the algorithm.  
Then increment the time (in mcs) by 
$\tau_i = \tau_i + \Delta\tau_i$ with 
\begin{equation}
\Delta\tau_i = 
\left\lfloor
\frac{\ln(r_1)}{\ln\left(1-\frac{p_{\rm sum}}{N_{\rm w}}\right)}
\right\rfloor + 1
\end{equation}
with $p_{\rm sum}=\sum_{k=1}^{N_{\rm w}} p_{\rm jump}(k)$ and 
$r_1$ a uniformly distributed random number.  

Next use a random number $r_2$ to pick one of the $j$ walkers 
to move, 
in the same fashion as was done for the $n$-fold way method, 
Eq.~\ref{jpick}.  
The only difference here is that when one of the $N_{s2}$ 
walkers is picked it will exit from the 5-6 or 16-17 minima, not just 
move to an adjoining lattice site.  

Next use random numbers to find the number of times $m_i$ that each 
of the $N_{\rm w}$ walkers was picked given that the system exited 
from the current state at time $\Delta\tau_i$.  
One way of doing this is with a tree-like structure.  
For example, for 
$N_{\rm w}=4$ use 3 random numbers $r_3$, $r_4$ and $r_5$, with the 
number of times walker 1 was picked equal to $m_1=r_4 r_3 (\Delta\tau_i-1)$, 
the number of times walker 2 was picked $m_2=(1-r_4) r_3 (\Delta\tau_i-1)$, 
for walker 3 one has $m_3=r_5 (1-r_3) (\Delta\tau_i-1)$, 
and for walker 4 one has $m_4=(1-r_5) (1-r_3) (\Delta\tau_i-1)$.  
Make sure that rounding effects does not change $\Delta\tau_i$, 
i.e.\ ensure that $\Delta\tau_i=1+\sum_{\ell=1}^{N_{\rm w}} m_\ell$.  
For each of the $N_{s2}$ walkers use its own random number $r_x$, 
and move walker $i$ to its other equal-energy 
minimum site if 
\begin{equation}
r_x\le \frac{1}{2} - 
\frac{1}{2}\left(\frac{1}{2}\right)^{m_i} , 
\label{s2rattle}
\end{equation}
otherwise leave it where 
it was.  

If the $j^{\rm th}$ walker that was picked to move is in 
the 5-6 or 16-17 minima, move it out from the minima to its 
adjacent higher-energy site.  If the $j^{\rm th}$ walker is 
not in one of these minima, then as in the $n$-fold way, 
use a random number 
$r_y$ and 
the probability of moving left, $p_{\rm move,-}$, from 
Eq.~(\ref{pmove}), and move the $j^{\rm th}$ walker left if 
\begin{equation}
p_{\rm move,-} \le r_y p_{\rm jump}(j), 
\end{equation}
otherwise it is moved 
to the right.  
This algorithm is repeated until all walkers coagulate at one point.  

As seen in the figures, with this algorithm denoted by {\it amc1\/}, 
the average lifetimes for this algorithm are also equal (within 
statistical errors for $M=10^3$ escapes) to the results from the 
other two algorithms.  However, as seen in Fig.~\ref{fig:f4}, the 
time required for the simulation at low temperatures is 
approximately independent of temperature!  Thus this algorithm is many 
times faster at low temperatures than even the $n$-fold way algorithm.  
At $T^{-1}=50$ this algorithm is about $10^8$ times faster than the 
$n$-fold way algorithm.  

The reason this algorithm scales so well at low temperatures is 
that in one algorithmic step one of the walkers moves to a site 
which has a higher energy, and consequently such a move will 
become less probable at low temperatures.  
The jump probability of 
Eq.~(\ref{s2rattle}) can be obtained using a Markov chain with 
the probability of remaining at the current lattice site 
equal to $\frac{3}{4}$ and the probability of going to the 
other equal-energy site equal to $\frac{1}{4}$.  Note that we already 
know this walker stays in its minima for $m_i$ time steps, which 
allows us to use a Markov chain.  

In general higher $s$ MCAMC algorithms could also be used.  For 
certain energy landscapes (for example where a walker will rattle 
around in a local energy minimum composed of more than two sites)
such higher $s$ MCAMC algorithms will be needed to obtain an algorithm 
that scales approximately independently of $T$ for low temperatures.  

\section{Results}

The results in Fig.~\ref{fig:f2}
and \ref{fig:f3} show that 
there is a minimum in $\langle\tau\rangle$ 
with temperature.  At low temperatures for large $N_{\rm w}$, 
most likely there is at least one walker that will be in the 
16-17 minimum, and this walker will require a long time to 
move over the saddle point (at site 12) to join the other 
walkers (which by that time will probably all be at the global minimum, 
site~9).  At high temperatures, the number of steps before 
all walkers will be at the same site should be approximately the 
same as the probability of them all being at the same site 
if they are placed randomly on the lattice, a 
probability given here by 
$\left(\frac{1}{20}\right)^{N_{\rm w}-1}$.  
For high temperatures this can be seen in Fig.~\ref{fig:f3}.  

Thus there is a temperature where the lifetime $\langle\tau\rangle$ 
is a minimum.  
Furthermore, the approximate width of this minimum decreases 
with $N_{\rm w}$, as seen in Fig.~\ref{fig:f3}.  
This should be expected in other models 
with many degrees of freedom 
that have 
complicated energy for free-energy landscapes, such as 
models for protein folding.  

\section{Conclusion and Discussion}

The Monte Carlo with Absorbing Markov Chains (MCAMC) algorithm was 
introduced and applied to a simple model.  This model could be realized 
experimentally (in a course-grained fashion) using nanoscale magnetic 
films with non-constant widths.  It can also be viewed as a toy model 
to study the average time in which a protein will fold, or any other 
such model with a complicated energy surface and intrinsic dynamics.  
The model shows that there is a minimum in the average lifetime 
$\langle\tau\rangle$ as a function of temperature.  At low temperature 
the lifetime is large because a large energy barrier must be overcome.  
At high temperatures the lifetime to coagulation is large because 
it is improbable that the walkers will remain in a low energy 
configuration very long.  

At low temperatures, the 
MCAMC algorithm gives exponentially fast speed-ups compared to the 
traditional kinetic Monte Carlo algorithm.  The $s$$=$$2$ MCAMC 
algorithm in fact scales almost independently of temperature at 
low temperatures, compared to an exponentially growing simulation 
time required for the traditional kinetic Monte Carlo and the 
$n$-fold way simulations.  Such faster-than-real-time algorithms 
are required for realistic dynamic simulations of many model systems.  

\begin{acknowledgments}
Useful discussions with Prof.\ P.A.\ Rikvold are acknowledged.  
This research is funded partly the the U.S.\ National Science 
Foundation through grant DMR-0120310.  
\end{acknowledgments}

\begin{figure}[htbp] 
\includegraphics[width=7.5cm]{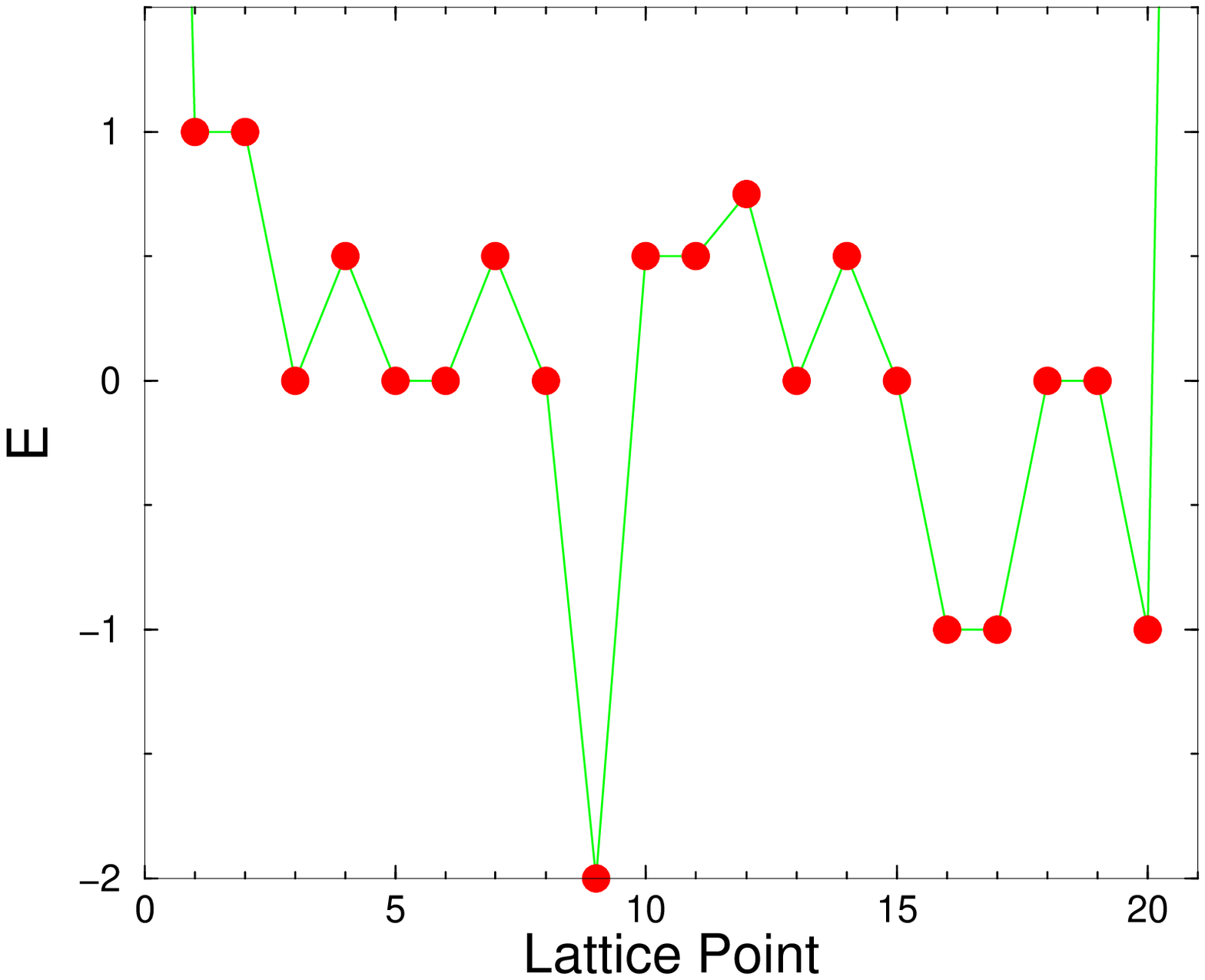} 
\caption{
\label{fig:EEE}
The energies of the 20 sites are shown, and are listed in Table~1.  
}
\end{figure}

\begin{figure}[htbp] 
\includegraphics[width=7.5cm]{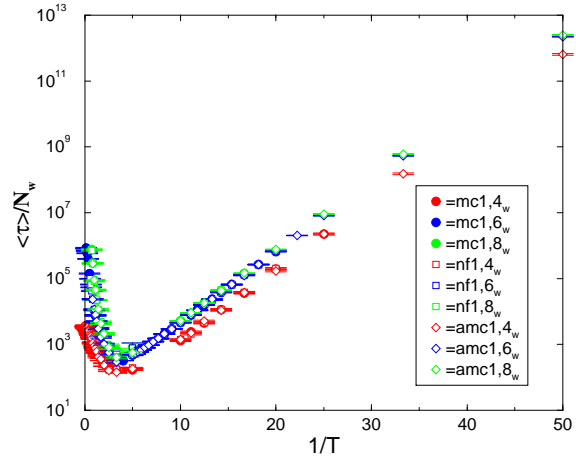}  
\caption{
The average lifetime per walker, 
$\langle\tau\rangle N^{-1}_{\rm w}$ 
from $10^3$ escapes 
is shown as a function of $T^{-1}$.  
Note the logarithmic scale.  
The solid lines join the points for 6 walkers.  
The 9 different points are for 3 different values of 
walkers ($N_{\rm w}=4,\>6,\>8$) and 3 different programs 
labeled {\it mc1\/} for normal Monte Carlo, {\it nf1\/}
for $n$-fold way, and {\it amc1\/} for $s$$=$$2$
MCAMC.  
\label{fig:f2}
}
\end{figure}

\begin{figure}[htbp] 
\includegraphics[width=7.5cm]{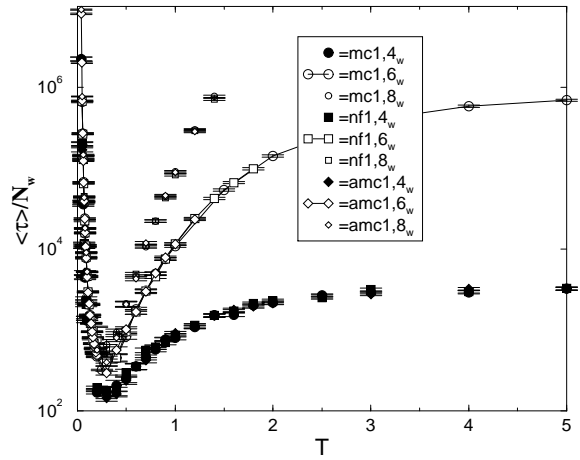}  
\caption{
The same as Fig.~2 but plotted as a function of $T$ to show the behavior at 
high temperatures.  All symbols and notation is the same as Fig.~2.
\label{fig:f3}
}
\end{figure}

\begin{figure}[htbp] 
\includegraphics{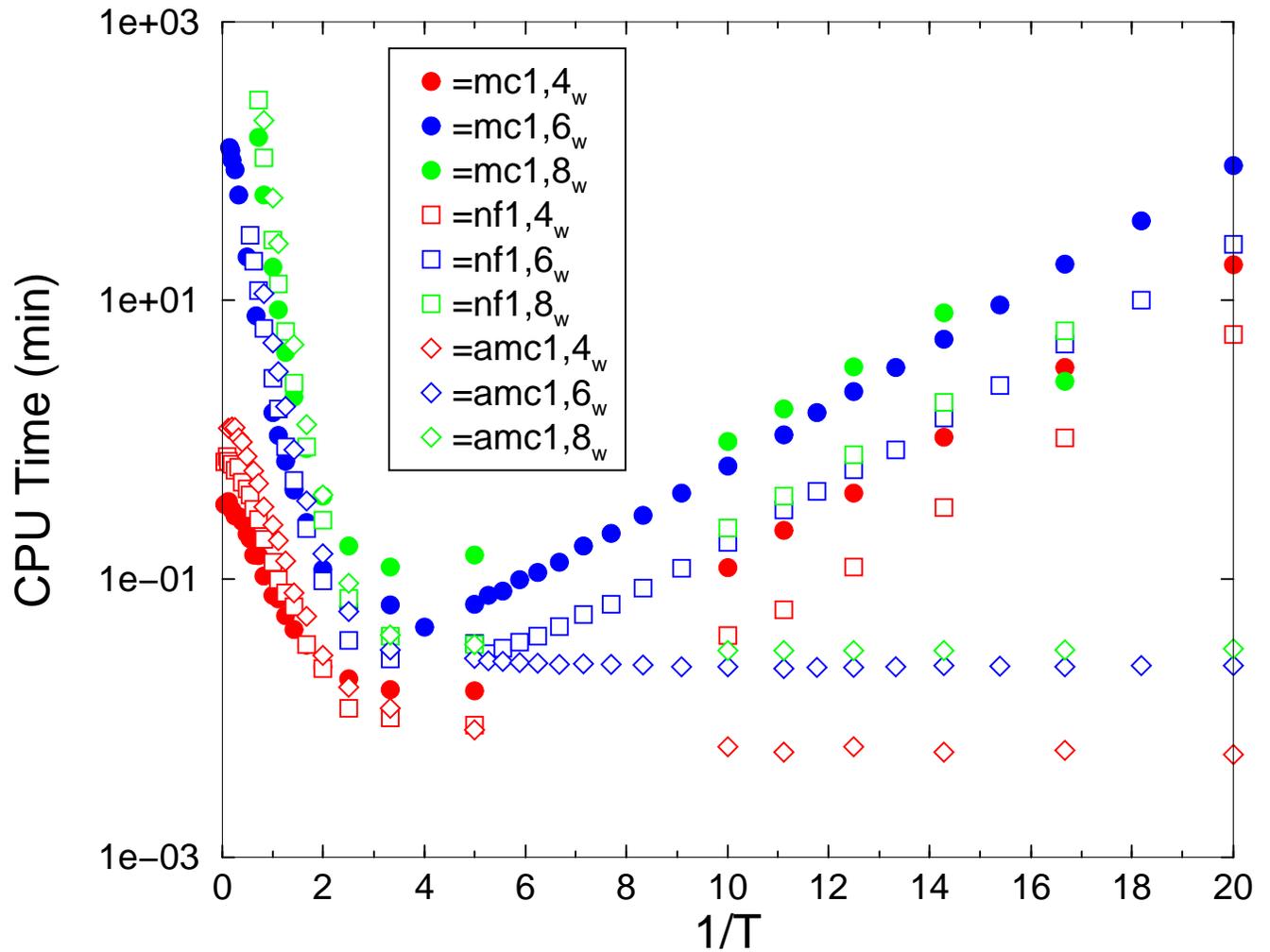}  
\caption{
The CPU time required in minutes to run $10^3$ escapes 
on a vector Cray SV1 computer is shown as a function of $T^{-1}$.  
Note the logarithmic time scale.  
The solid lines join the points for 6 walkers.  
The 9 different points are for 3 different values of 
walkers ($N_{\rm w}=4,\>6,\>8$) and 3 different programs 
labeled {\it mc1\/} for normal Monte Carlo, {\it nf1\/}
for $n$-fold way, and {\it amc1\/} for $s$$=$$2$
MCAMC.  
\label{fig:f4}
}
\end{figure}

\end{document}